\documentclass[twocolumn,showpacs,aps,floatfix]{revtex4}
\usepackage{graphicx}
\usepackage{amsmath2000}
\begin{document}
\title{Energies, transition rates, and electron electric dipole moment
enhancement factors for  Ce IV and Pr V}
\author{I. M. Savukov}
 \email{isavukov@princeton.edu}
 \homepage{http://www.nd.edu/~isavukov}
 \author{W. R. Johnson}
 \email{johnson@nd.edu}
 \homepage{http://www.nd.edu/~johnson}
 \author{U. I. Safronova}
 \email{usafrono@nd.edu}
\affiliation{Department of Physics, 225 Nieuwland Science Hall\\
University of Notre Dame, Notre Dame, IN 46566}
 \author{M. S. Safronova}
 \email{msafrono@nd.edu}
\affiliation{Electron and Optical Physics Division,
National Institute for Standards and Technology\\
Gaithersburg, MD, 20899}
\date{\today}
\begin{abstract}
Energies, transition rates, and electron electric dipole moment (EDM) 
enhancement factors are calculated for low-lying states of
Ce IV and Pr V using relativistic many-body perturbation theory. 
This study is related to recent investigations of the more complicated 
Gd IV ion, which is promising for electron EDM experiments. 
The ions Ce IV and Pr V both have a single valence electron, permitting 
one to carry out reliable {\it ab-initio} calculations
of energy levels, transition rates and other atomic properties
using well developed computational methods.
\end{abstract}
\maketitle

\section{Introduction}

Recently, searches for an electron electric dipole moment (EDM) 
in atoms and molecules have gained considerable interest. 
Since an electron EDM violates
time-reversal symmetry and parity, 
the discovery of an EDM would have many
implications for modern fundamental theories 
(a general overview is given by \citet{KL}). 
The best limit on the value of the electron EDM was obtained 
by \citet{Com} in Tl, $d_{e}<1.6\times 10^{-27}$ e cm. 
Although there is a large enhancement ($\sim 600$) of the electron EDM
in Tl, the density of atoms in a beam is much
lower than in a solid-state system.
To utilize the advantage of high densities, 
it has been proposed to use Gadolinium Gallium Garnet 
Gd$_{3}$Ga$_{5}$O$_{12}$ or Gadolinium Iron Garnet
Gd$_{3}$Fe$_{5}$O$_{12}$  (densities $\sim 10^{22}$/cc)
in EDM experiments \cite{lam,Hun}.
There are various features of these materials that make them 
particularly useful for such experiments. For example, 
magneto-electric effects are forbidden owing to the FCC
symmetry of the crystals, simplifying the exclusion of systematics;
the crystal Gd$_{3}$Fe$_{5}$O$_{12}$ has a very high
resistivity ($\gg 10^{16}\;\Omega$-cm); and spin alignment is
relatively easy. The electron EDM enhancement for the 
Gd$^{3+}$ ion ($\sim 2-3$)
is much smaller than for Tl; nevertheless there is a substantial 
gain in the number of atoms.

Motivated by proposed EDM measurements, calculations of EDM enhancement 
have been
performed recently \cite{Buh,Kuen}. Core polarization effects in Gd IV were
discussed by \citet{Dzuba} and found to decrease the size of EDM
enhancement factor from -3.3 to -2.2 . Thus, there is a strong sensitivity 
of the EDM enhancement to many-body effects and further investigation is necessary.

The ion Gd$^{+3}$, which has a $4f^{7}$ ground-state configuration 
is extremely complicated and difficult for atomic theory,
though some understanding was gained by \citet{Dzuba} using a relativistic 
configuration-interaction (RCI) method
and the widely used Cowan~\cite{cowan} code. As mentioned above, 
core polarization 
was found to be substantial. For example, to match 
the experimental $4f\rightarrow 5d$
transition energies,   scaling factors of 0.8 and 0.85
for Coulomb integrals were used in the RCI and Cowan codes, respectively.  
In addition, a polarization potential was introduced in RCI code to match
experimental energies and a systematic energy shift of $\sim 18,000$ cm$^{-1}$
was made in the Cowan code. This shift can be attributed to 
core polarization by a single valence electron. An {\it ab-initio} investigation of
correlation should help to understand these empirical adjustments.
The simplest ions (those having a single $4f$ valence electron) which 
exhibit core-polarization effects similar to Gd IV are Ce IV and Pr V. 
To gain a clearer understanding of the role of core polarization and other
many-body effects in Ce IV and Pr V, we calculate energies, 
dipole transition matrix elements,
and EDM enhancement factors for low lying states using 
relativistic many-body perturbation theory (MBPT). 

The present calculations of energies
are carried out to third order in MBPT using the methods developed earlier to study the
Li, Na, and Cu isoelectronic sequences \cite{liseq,naseq, cuseq}.
The calculations of transition matrix elements is also carried out to third order 
in MBPT using the methods developed in \cite{alktran} to treat the alkali atoms and
alkalilike ions.  Here, we use the gauge-independent  
version of the third-order MBPT code described in \cite{gitran}. 
The calculations of electron EDM enhancement factors, which involve a sum over
intermediate states, are carried out in the random-phase approximation (RPA)
following a procedure similar to that described in \cite{weak2}.

It should be mentioned that Ce$^{+3}$, which has a $4f_{5/2}$ ground state
and an observed magnetic moment 2.3-2.5 $\mu_B$ can be embedded into garnet 
crystals and used in EDM experiments of the type proposed in \cite{lam}.
In those experiments, 
a strong electric field polarizes ions which in turn produce a small magnetic
field that is measured in a sensitive SQUID detector. The magnetic field
at low temperatures is proportional to the product 
of the electric dipole moment of the ion and its magnetic moment. 
The estimated EDM of Gd$^{+3}$
as is a factor of about 3 larger than for Ce$^{+3}$ 
and the magnetic moment of Gd$^{+3}$ (7.9-8.0 $\mu_B$) 
is also about three times larger; therefore,
there is an overall advantage of nine in Gd$^{+3}$ compared
to Ce$^{+3}$.
Nevertheless, the Ce$^{+3}$ ion could still compete in setting experimental
limits on the electron EDM owing to the fact that its ionic EDM 
has significantly smaller theoretical uncertainty.

\section{Calculation of energies}

First-, second- and third-order Coulomb energies $E^{(n)},\ n=1..3$,  and
first- and second-order
Breit energies $B^{(n)},\ n=1,2$, calculated using methods described in
Refs.~\cite{liseq,naseq,cuseq}, are presented in Table~\ref{taben} along
with the resultant theoretical energies $E_\text{tot}$ and predicted
energies $E_\text{NIST}$ from the 
National Institute of Standards and Technology (NIST) 
given by \citet{martin}.  We see that 
second-order corrections are large and improve the accuracy
of the first-order Dirac-Hartree-Fock (DHF) energies. Third-order MBPT 
further improves the ground state energy. However,
the third-order correction is relatively large (roughly one-third of the second
order) and overshoots the experimental value, which indicates that
oscillations in higher orders are likely. Using a geometric progression with 
$q=-1/3$ we can extrapolate the second- and third-order values to 
give a limiting ground-state energy $-296420$ cm$^{-1}$ for Ce$^{+3}$,
in close agreement with the predicted value from \cite{martin}. 
The corresponding 
extrapolation for Pr$^{+4}$ gives $-468290$ cm$^{-1}$ differing from the
predicted ground-state energy given in \cite{martin} by about 4000 cm$^{-1}$.
This rather large difference
casts doubt on the threshold energy of Pr V predicted in \cite{martin} and  
explains the
large differences with the NIST energies seen in 
lower half of Table~\ref{taben}. 

In Refs.~\cite{liseq,naseq,cuseq}, where the ionic ground-states were $2s$, $3s$,
and $4s$, respectively, the third-order correction was uniformly much smaller
than the second-order correction. The relatively large size of the third-order
corrections in Ce IV and Pr V arise because of the double-well feature of the $4f$ Coulomb potential
discussed, for example, by \citet{KTFF:83}.  

\begin{table}
\caption{\label{tab1} First-order (DHF) energies $E^{(1)}$, 
second- and third-order Coulomb energies $E^{(2)}$ and $E^{(3)}$,
first- and second-order Breit corrections $B^{(1)}$ and $B^{(2)}$
and totals $E_\text{tot}$ for Ce IV and Pr V
are compared with predicted energies $E_\text{NIST}$ given by 
\protect\citet{martin}. 
Units: cm$^{-1}$   \label{taben}}
\begin{ruledtabular}
\begin{tabular}{lrrrrrrrr}
\multicolumn{1}{c}{$nlj$ } &
\multicolumn{1}{c}{$E^{(1)}$} &
\multicolumn{1}{c}{$B^{(1)}$} &
\multicolumn{1}{c}{$E^{(2)}$} & 
\multicolumn{1}{c}{$B^{(2)}$} &
\multicolumn{1}{c}{$E^{(3)}$} &
\multicolumn{1}{c}{$E_\text{tot}$} &
\multicolumn{1}{c}{$E_\text{NIST}$} &
\multicolumn{1}{c}{$\delta E$} \\
\hline
\multicolumn{8}{c}{Ce~IV}\\
\hline
$4f_{5/2}$& -261361&  616& -46747&  -2552& 15672&  -294372& -296470&  2098\\
$4f_{7/2}$& -259378&  423& -46091&  -2511& 15398&  -292158& -294217&  2059\\
$5d_{3/2}$& -236793&  292& -13516&   -545&  3753&  -246809& -246733&   -76\\
$5d_{5/2}$& -234637&  220& -12957&   -528&  3675&  -244226& -244244&    18\\
$6s_{1/2}$& -203245&  168& -10201&   -200&  3680&  -209794& -209868&    70\\
$6p_{1/2}$& -168978&  167&  -7283&   -127&  2329&  -173891& -173885&    -6\\
$6p_{3/2}$& -164703&  120&  -6648&   -122&  2118&  -169236& -169178&   -58\\
$6d_{3/2}$& -116268&   67&  -3989&   -106&   997&  -119300& -119272&   -28\\
$6d_{5/2}$& -115581&   51&  -3922&   -106&   977&  -118580& -117557& -1024\\
$7s_{1/2}$& -110878&   65&  -3601&    -74&  1293&  -113196& -112968&  -226\\
\hline
\multicolumn{8}{c}{Pr~V}\\
\hline
$4f_{5/2}$& -431686&  798& -48799&  -3008& 16527&  -466167& -464000& -2167\\
$4f_{7/2}$& -428863&  551& -48175&  -2965& 16270&  -463182& -460973& -2209\\
$5d_{3/2}$& -341470&  395& -14968&   -669&  4088&  -352624& -348948& -3676\\
$5d_{5/2}$& -338314&  298& -14386&   -651&  3827&  -349225& -345486& -3739\\
$6s_{1/2}$& -281133&  225& -11649&   -247&  4241&  -288563& -285029& -3535\\
$6p_{1/2}$& -239154&  239&  -9136&   -169&  2451&  -245769& -240522& -5247\\
$6p_{3/2}$& -233035&  171&  -8489&   -163&  2040&  -239477& -233961& -5516\\
$6d_{3/2}$& -171610&   98&  -5746&   -141&   792&  -176576&        &	  \\
$6d_{5/2}$& -170572&   75&  -5647&   -141&   771&  -175481&        &	  \\
$7s_{1/2}$& -156266&   92&  -4489&    -97&  1329&  -161431& -159489& -1942\\
\end{tabular}
\end{ruledtabular}
\end{table}

\section{Calculation of transition matrix elements and transition rates}

Transition matrix elements provide another test of quality of atomic-structure
calculations and another measure of the size of correlation corrections. 
Third-order MBPT reduced matrix elements for 
transitions between low-lying states of Ce$^{+3}$ and Pr$^{+4}$ 
are presented in Table~\ref{tran}.  The first-order reduced matrix elements
$Z^{(1)}$  are obtained from length-form DHF calculations. Length-form and 
velocity-form matrix elements differ typically by 10\%. Second-order
matrix elements in the table $Z^{(2)}$, which include $Z^{(1)}$, are extended 
to include all higher-order corrections
associated with the random-phase approximation. These second-order
calculations are practically gauge independent. In the present calculations, 
length- and velocity-form
matrix elements in the RPA agree to six or more digits. The third-order
matrix elements $Z^{(3)}$ include $Z^{(2)}$ plus Brueckner-orbital (BO),
structural radiation, and normalization corrections described, for example,
in \cite{alktran}.  These calculations are carried out in a gauge-independent
manner, including appropriate derivative terms, as described in \cite{gitran}.
We truncated our basis set to include only those partial waves with 
$l\le 8$, and found that length- and velocity-form third-order 
reduced matrix elements  agreed to 4 digits.

As can be seen in Table~\ref{tran}, RPA corrections are very large, 10-40\%, 
being largest for 
$4f\rightarrow 5d$ transitions, and must be taken into account. Such behavior
can be attributed to core shielding which is substantial because valence
electrons penetrate deeply into the core. Third-order corrections are
smaller, 2-4\%  scaling as $1/Z_\text{ion}$.  If such scaling holds in
higher orders, we can estimate the accuracy of our calculations to be
0.4-0.8\%. The dominant contribution in third order comes from
the BO correction which is approximately equal to the sum of
the other third-order corrections.

\begin{table}
\caption{Reduced matrix elements of the dipole operator in first-, second-, and
third-order perturbation theory for transitions
in Ce IV and Pr V.\label{tran}}
\begin{ruledtabular}
\begin{tabular}{ccccccc}
\multicolumn{1}{c}{} &
\multicolumn{3}{c}{Ce IV} &
\multicolumn{3}{c}{Pr V} \\
\multicolumn{1}{c}{Transition} &
\multicolumn{1}{c}{Z$^{(1)}$} &
\multicolumn{1}{c}{Z$^{(2)}$} &
\multicolumn{1}{c}{Z$^{(3)}$} &
\multicolumn{1}{c}{Z$^{(1)}$} &
\multicolumn{1}{c}{Z$^{(2)}$} &
\multicolumn{1}{c}{Z$^{(3)}$} \\
\hline
$4f_{5/2}\rightarrow 5d_{3/2}$ & 1.498& 0.972& 1.172& 1.146& 0.678& 0.706\\
$4f_{5/2}\rightarrow 5d_{5/2}$ & 0.396& 0.264& 0.308& 0.302& 0.186& 0.189\\
$4f_{7/2}\rightarrow 5d_{5/2}$ & 1.799& 1.193& 1.413& 1.370& 0.828& 0.852\\
$5d_{3/2}\rightarrow 6p_{1/2}$ & 1.976& 1.768& 1.682& 1.648& 1.471& 1.396\\
$5d_{3/2}\rightarrow 6p_{3/2}$ & 8.363& 0.756& 0.719& 0.690& 0.625& 0.592\\
$5d_{5/2}\rightarrow 6p_{3/2}$ & 2.585& 2.352& 2.229& 2.140& 1.950& 1.837\\
$6s_{1/2}\rightarrow 6p_{1/2}$ & 2.847& 2.482& 2.402& 2.560& 2.207& 2.136\\
$6s_{1/2}\rightarrow 6p_{3/2}$ & 4.012& 3.512& 3.401& 3.609& 3.125& 3.020\\
\end{tabular}
\end{ruledtabular}
\end{table}

Transition rates $A$ (s$^{-1}$), oscillator strengths $f$, and wavelengths
$\lambda$ ({\AA}) for electric dipole transitions between low-lying
states of Ce IV and Pr V
are given in Table~\ref{comp}.
These data are calculated using the dipole matrix elements 
$Z^{(3)}$ from Table~\ref{tran} and predicted
NIST transition energies \cite{martin}. In the two final columns of Table~\ref{comp},
we compare our MBPT wavelengths with the wavelengths from Ref.~ \cite{martin}.
We also compare our MBPT oscillator strengths with theoretical oscillator strengths obtained
by \citet{migdalek}.
The data in \cite{migdalek} were obtained using a relativistic model potential (RMP)
approach together with a core-polarization (CP) model potential.
Our data and that from Ref.~\cite{migdalek} agree well for $5d-6p$ and $6s-6p$
transitions but differ for $4f-5d$ transition where $f$ values are very small.

\begin{table}
\caption{\label{comp} MBPT transition rates $A$ (s$^{-1}$),
oscillator strengths $f$,  and wavelengths $\lambda$ (\AA) for
transitions in Ce IV and Pr V. MBPT (a) oscillator
strengths are compared with theoretical calculations (b) performed
in Ref.~\protect\cite{migdalek}. MBPT (a) wavelengths are compared with
wavelengths (c) predicted by NIST~\protect\cite{martin}.}
\begin{ruledtabular}
\begin{tabular}{llllrr}
\multicolumn{1}{c}{Transition} &
\multicolumn{1}{c}{$A^{(a)}$} &
\multicolumn{1}{c}{$f^{(a)}$} &
\multicolumn{1}{c}{$f^{(b)}$} &
\multicolumn{1}{c}{$\lambda^{(a)}$} &
\multicolumn{1}{c}{$\lambda^{(c)}$}\\
\hline
\multicolumn{6}{c}{Ce IV} \\
\hline
$4f_{5/2}\rightarrow 5d_{3/2}$& 8.56[7]& 0.0346 &0.0159 & 2102 & 2011\\
$4f_{5/2}\rightarrow 5d_{5/2}$& 4.56[6]& 0.0025 &0.0013 & 1994 & 1915\\
$4f_{7/2}\rightarrow 5d_{5/2}$& 8.41[7]& 0.0379 &0.0186 & 2086 & 2001\\
$5d_{3/2}\rightarrow 6p_{1/2}$& 1.11[9]& 0.157  &0.158  & 1371 & 1373\\
$5d_{3/2}\rightarrow 6p_{3/2}$& 1.22[8]& 0.0304 &0.0206 & 1289 & 1289\\
$5d_{5/2}\rightarrow 6p_{3/2}$& 1.06[9]& 0.189  &0.189  & 1334 & 1332\\
$6s_{1/2}\rightarrow 6p_{1/2}$& 2.72[8]& 0.315  &0.327  & 2785 & 2779\\
$6s_{1/2}\rightarrow 6p_{3/2}$& 3.95[8]& 0.715  &0.745  & 2465 & 2458\\
\hline
\multicolumn{6}{c}{Pr V} \\
\hline
$4f_{5/2}\rightarrow 5d_{3/2}$& 3.85[8]&0.0290 &0.0318 &  881 & 869 \\
$4f_{5/2}\rightarrow 5d_{5/2}$& 2.01[7]&0.0021 &0.0019 &  863 & 844 \\
$4f_{7/2}\rightarrow 5d_{5/2}$& 3.78[8]&0.0318 &0.0285 &  878 & 866 \\
$5d_{3/2}\rightarrow 6p_{1/2}$& 2.52[9]&0.161  &0.165  &  936 & 922 \\
$5d_{3/2}\rightarrow 6p_{3/2}$& 2.70[8]&0.0306 &0.0311 &  884 & 870 \\
$5d_{5/2}\rightarrow 6p_{3/2}$& 2.37[9]&0.191  &0.193  &  911 & 897 \\
$6s_{1/2}\rightarrow 6p_{1/2}$& 4.07[8]&0.308  &0.321  &  2337& 2247 \\
$6s_{1/2}\rightarrow 6p_{3/2}$& 6.15[9]&0.707  &0.746  &  2037& 1958 \\
\end{tabular}
\end{ruledtabular}
\end{table}

\section{EDM enhancement}

\subsection{Basic equations}

According to Schiff's theorem \cite{schiff:63}, the electric dipole moment of an atom
induced by an intrinsic electron EDM vanishes in the nonrelativistic limit;
however, as shown by \citet{sandars:68}, the atomic EDM is nonvanishing relativistically
and can be a large multiple of the intrinsic electron moment for heavy atoms.   
If we assume that the electron has an intrinsic EDM $d_e$, 
then the EDM of a many-electric atom $D$ 
may be written \cite{weak2}
\begin{equation}
 D = 2  \sum_n \frac{\langle v j_v | e Z | n \rangle \
\langle n | H_\text{edm} \ | v j_v \rangle}{E_v - E_n} ,
\end{equation}
where $eZ$ is the dipole electric operator
\[
eZ = \sum_i e z_i ,
\] 
and $H_\text{edm}$ is an equivalent EDM interaction \cite{weak2} given by
\[
 H_\text{edm} = -2i\frac{d_e}{e}\, c \, \sum_j p_j^2\, \beta_j\, (\gamma_5)_j = H^\dagger_\text{edm} \, .
\]
This equivalent interaction, which automatically accounts for Schiff's theorem,
is rotationally invariant and therefore
conserves angular momentum; it violates both parity and time-reversal symmetry.

For an atom or ion with one valence electron,
one-electron matrix elements of $H_\text{edm}$ may be written in lowest order as
\begin{equation}
\langle n m_n | H_\text{edm} | v j_v \rangle =  \delta_{\kappa_n\, -\kappa_v}
\delta_{m_n j_v} \ \langle n \| H_\text{edm} \| v \rangle ,
\end{equation}
where the (somewhat unconventional) reduced matrix element is
\begin{multline}
\langle n \| H_\text{edm} \| v \rangle =  \nonumber \\
 2\, c\, \frac{d_e}{e}\, \int_0^\infty\!\!\!  dr  \left[ G_n(r) \frac{d^2 F_v}{dr^2}
 - \frac{\kappa_v(\kappa_v-1)}{r^2} G_n(r) F_v(r)  \right. \nonumber \\
 \left.\hspace{2em} +  F_n(r) \frac{d^2 G_v}{dr^2}
- \frac{\kappa_v(\kappa_v+1)}{r^2} F_n(r) G_v(r) \right] .
\end{multline}
In the above equation, $G_k(r)$ and $F_k(r)$ are the large and small components, respectively,
of radial Dirac wave functions.
Similarly, we may write
\begin{equation}
\langle v j_v | eZ | n j_v \rangle = \sqrt{\frac{j_v}{(2j_v+1)(j_v+1)}}\ \langle v \|
eZ \| n \rangle
\end{equation}
with
\begin{multline}
\langle v \| eZ \| n \rangle = e\ \langle \kappa_v \| C_1 \| -\kappa_v \rangle \times \nonumber\\
\int_0^\infty \!\! r dr \left[ G_v(r)G_n(r) + F_v(r) F_n(r) \right] ,
\end{multline}
$C_{1q}(\hat{r})$ being a normalized spherical harmonic.
The expression for the atomic dipole moment in lowest-order MBPT then reduces to 
\begin{equation}
D^{(1)} = 2\ \sqrt{\frac{j_v}{(2j_v+1)(j_v+1)}}\ \sum_i 
\frac{\langle v \| eZ \| i \rangle\ \langle i \| H_\text{edm} \| v \rangle}
{\epsilon_v - \epsilon_i} , \label{eqbare}
\end{equation}
where $\epsilon_k$ are eigenvalues of the valence-electron Dirac equation.

\subsection{RPA correlation corrections}

Lowest-order calculations of the induced atomic EDM are carried out in a frozen-core
 $V^{N-1}$ DHF potential. Such calculations were shown in \cite{weak2}
to be very sensitive to correlation corrections. For that reason, the lowest-order ``bare''
matrix elements in Eq.~(\ref{eqbare}) are replaced by ``dressed'' RPA matrix elements.

\subsubsection{Z-RPA}
 Thus,  we replace the lowest-order dipole matrix element
$\langle w || Z || v \rangle$ in Eq.~(\ref{eqbare})
by
\begin{eqnarray}
\lefteqn{\langle w || Z^\text{RPA} || v \rangle = \langle w || Z || v \rangle}\hspace{0em}\nonumber\\
&& +\sum_{an} (-1)^{a-n+1} \frac{1}{3} 
\frac{\langle a || Z^\text{RPA} || n \rangle Z_1(wnva)}
{\epsilon_a - \epsilon_n} \nonumber \\
&&+ \sum_{an} (-1)^{a-n+1} \frac{1}{3} 
\frac{Z_1(wavn) \langle n || Z^\text{RPA} || a \rangle}
{\epsilon_a - \epsilon_n} \, , \label{eq2}
\end{eqnarray}
where the index $a$ extends over all core orbitals and the index $n$ extends
over all virtual orbitals permitted by angular-momentum selection rules.
The quantities $Z_J(ijkl)$ are Coulomb integrals
\begin{equation}
Z_J(ijkl) =  X_J(ijkl) + [J] \sum_L 
\left\{
\begin{array}{ccc}
i  &  k  & J \\
l  &  j  & L 
\end{array}
\right\}
X_L(ijlk) \ ,
\end{equation}
where $X_J(ijkl)$ are defined by
\[
X_J(ijkl) = (-1)^J \langle i \| C_J \| k  \rangle\ \langle j \| C_J \| l  \rangle\
R_J(ijkl) \, ,
\]
$R_J(ijkl)$ being a Slater integral
\begin{multline*}
R_J(ijkl) = \int_0^\infty\!\!\! \int_0^\infty\!\!\!  drdr' \frac{r_<^J}{r_>^{J+1}}
\left[G_i(r)G_k(r)+F_i(r)F_k(r)\right] \\
\left[G_j(r')G_l(r')+F_j(r')F_l(r')\right] .
\end{multline*}

 We designate the corresponding approximation to the
atomic EDM by $D^\text{RPA}_Z$.  
Note that if we replace $\langle n || Z^\text{RPA} || a \rangle$ by
$\langle n || Z || a \rangle$ on the right hand side of Eq.~(\ref{eq2}), then
we obtain the second-order correlation correction to the valence-excited dipole matrix element.
The atomic EDM calculated in this approximation is designated by $D^{(2)}_Z$.

\subsubsection{H-RPA}
Similarly, we replace the bare matrix element of the EDM interaction
$\langle w \| H_\text{edm} \| v \rangle$ by its dressed counterpart
\begin{eqnarray}
\lefteqn{\langle w || H^\text{RPA}_\text{edm} || v \rangle = 
\langle w || H_\text{edm}  || v \rangle}\hspace{0em}\nonumber\\
&& + \sum_{an} \sqrt{\frac{[j_a]}{[j_v]}} 
\frac{\langle a || H^\text{RPA}_\text{edm} || n \rangle Z_0(wnva)}
{\epsilon_a - \epsilon_n} \nonumber \\
&&+ \sum_{an} \sqrt{\frac{[j_a]}{[j_v]}} 
\frac{Z_0(wavn) \langle n || H^\text{RPA}_\text{edm} || a \rangle}
{\epsilon_a - \epsilon_n} \, . \label{eq3}
\end{eqnarray}
We designate the approximation to $D$ obtained using the dressed matrix
element from Eq.~(\ref{eq3}) by $D^\text{RPA}_H$. Again, if we replace dressed
matrix elements by bare matrix elements on the right hand side of Eq.~(\ref{eq3}),
we obtain a second-order approximation to $\langle w || H_\text{edm} || v \rangle$.
The resulting correction to the atomic EDM is designated by $D^{(2)}_H$. 
 
Core-excited matrix elements $\langle n || Z^\text{RPA} || a \rangle$  and 
$\langle n || H^\text{RPA}_\text{edm} || a \rangle$ in Eqs.~(\ref{eq2}) and (\ref{eq3})
satisfy sets of coupled equations given explicitly in \cite{alktran}. 

\subsection{Calculations of EDM enhancement factors}
 
The sums over intermediate states in Eqs.~(\ref{eqbare}-\ref{eq3}) 
are carried out using basis functions
obtained as linear combinations of B-splines as described in \cite{bspline:88}.
We use 40 splines of order 7 and constrain the ions to lie in a cavity 
of radius $R=35$ a.u.\ for Ce$^{+3}$ and
30 a.u.\ for Pr$^{+4}$.  

A detailed breakdown of the contributions to $D$ for $4f$ states of
Ce IV is given in Table~\ref{tappr}, where we
list the DHF approximation, $D^{(1)}$, the second-order correction $D^{(2)}$,
the RPA approximation, $D^\text{RPA}$, and the individual contributions to 
the second-order and RPA corrections from the dipole and weak-interaction matrix elements.
One can see from the table that the correlation corrections to the weak-interaction
matrix element are comparable to or larger than the lowest order matrix element. 
Moreover, there are significant changes in these correlation corrections going from
second-order MBPT to full RPA calculations.

Finally, in Table~\ref{tedm}, we present DHF and RPA values of the EDM enhancement factors
$D/d_e$ for the low-lying $4f$, $5d$, $6s$, and $6p$ states of Ce IV and Pr V.

\begin{table}
\caption{Comparison of first-order, second-order, and RPA calculations of the
atomic EDM enhancement factor $D/d_e$ for $4f$ states of Ce IV.
 \label{tappr}}
\begin{ruledtabular}
\begin{tabular}{cccccccc}
\multicolumn{1}{c}{ state   }  &
\multicolumn{1}{c}{$D^{(1)}$}  &
\multicolumn{1}{c}{$\Delta D^{(2)}_H$}&
\multicolumn{1}{c}{$\Delta D^{(2)}_Z$}&
\multicolumn{1}{c}{$D^{(2)}$} &
\multicolumn{1}{c}{$\Delta D^{\text{RPA}}_H$}&
\multicolumn{1}{c}{$\Delta D^{\text{RPA}}_Z$}&
\multicolumn{1}{c}{$D^{\text{RPA}}$} \\
\hline
$4f_{5/2}$&   -0.382& -0.388&  0.332& -0.438& -0.785& 0.387& -0.780\\
$4f_{7/2}$&   -0.002& -0.033&  0.022& -0.013& -0.045& 0.015& -0.032\\
\end{tabular}
\end{ruledtabular}
\end{table}

\begin{table}
\caption{EDM enhancement factors $D/d_e$ for low-lying states of Ce IV and Pr V.
\label{tedm}} 
\begin{ruledtabular}
\begin{tabular}{ccccc}
\multicolumn{1}{c}{State} &
\multicolumn{1}{c}{$D^{(1)}$} &
\multicolumn{1}{c}{$\Delta D^{\text{RPA}}_H$} &
\multicolumn{1}{c}{$\Delta D^{\text{RPA}}_Z$} &
\multicolumn{1}{c}{$D^{\text{RPA}}$} \\
\hline
\multicolumn{5}{c}{Ce IV} \\
\hline
   $4f_{5/2}$&   -0.382&  -0.785&   0.387&  -0.780\\
   $4f_{7/2}$& -0.00225& -0.0451&  0.0151& -0.0323\\
   $5d_{3/2}$&    -1.95&   -3.38&   0.779&   -4.55\\
   $5d_{5/2}$&    0.425&  -0.347&  -0.136& -0.0628\\
   $6s_{1/2}$&     120.&    27.5&   -19.8&    128.\\
   $6p_{1/2}$&    -158.&   -30.9&    19.8&   -169.\\
   $6p_{3/2}$&     2.89&    7.24&   -1.07&    9.06\\
\hline
\multicolumn{5}{c}{Pr V} \\
\hline
   $4f_{5/2}$&   -0.142& -0.0806&   0.0926&  -0.130\\
   $4f_{7/2}$& -0.00266&-0.00444&-0.000912&-0.00802\\
   $5d_{3/2}$&    -1.80&   -2.91&    0.655&   -4.05\\
   $5d_{5/2}$&    0.174&  -0.827&    0.125&  -0.528\\
   $6s_{1/2}$&     127.&    27.3&    -22.2&    132.\\
   $6p_{1/2}$&    -157.&   -30.0&     21.7&   -166.\\
   $6p_{3/2}$&     2.78&    6.06&    -1.01&    7.83\\
\end{tabular}
\end{ruledtabular}
\end{table}
\section{Conclusion}

We have studied energies, transition probabilities, and EDM enhancement
factors for Ce$^{+3}$ and Pr$^{+4}$. 
We found that perturbation theory converges quite
slowly and that RPA
corrections are the dominant correlation corrections for transitions.
We use our third-order MBPT with
``dressed'' matrix elements to obtain accurate transition rates. 
The most interesting discovery is that RPA corrections modify 
lowest-order values of the EDM enhancement factor significantly.

\begin{acknowledgments}
The authors wish to thank M. Romalis for helpful suggestions on this
paper. The work of I.S. and  W.R.J. and was supported in part by National
Science Foundation Grant No.\ PHY-01-39928. U.I.S. acknowledges
support by Grant No.\ B516165 from Lawrence Livermore
National Laboratory. 

\end{acknowledgments}

\end{document}